\documentclass{elsart}
\newcommand{\eps}{\epsilon}
\newcommand{\Pint}{P_{ i}}

\usepackage{epsfig}

\begin{document}

\begin{frontmatter}

\begin{abstract}

We introduce an analytical model for population dynamics with intra-specific 
competition, mutation and 
assortative mating as basic ingredients. The set of equations 
that describes the time evolution of population size in a mean-field 
approximation may be decoupled. We find a phase transition leading to sympatric 
speciation as a parameter that quantifies competition strength is varied. 
This transition, previously found in a computational model, occurs to be of 
first order.
 
\end{abstract}

\title{Phase transition in a mean-field model for sympatric speciation}
\author{V. Schw\"ammle, K.~Luz-Burgoa, J. S. S\'a Martins and S.~Moss de Oliveira}
\address{Instituto de F\'{\i}sica, Universidade Federal Fluminense, 
Campus da Praia Vermelha, Boa Viagem, Niter\'oi, 24210-340, RJ, Brazil}

\end{frontmatter}

\section{Introduction}
The dynamics that generate the rich and diverse structure of the living world 
is still one of the greatest puzzles in science. Why and how the primitive 
living organisms gave birth to the immense variety of species in our 
environment is still a matter of debate and research to our days. The theory 
of biological evolution, the paradigm that guided the rapid growth of our 
knowledge about the microscopics (e.g. the behaviour of individuals in a 
population) of life and the development of all the dazzling techniques and 
possibilities of genetical engineering, has yet to answer some simple and 
rather basic questions. One of these stands out as a crucial difficulty: the 
issue of sympatric speciation. 

If a single-species population is somehow split into two separate groups - by 
the establishment of some geographical barrier, say - uncorrelated genetic 
drift in these non-mating populations may eventually lead to differentiation. 
This process is called allopatric speciation and is reasonably well understood. 
The same cannot be said about the branching of a single population into two 
distinct species without the appearance of an external barrier dividing the 
original group: sympatric speciation. Until recently, the possibility of such a 
process was still under debate, but observations of micro-evolution 
\cite{friesen} and the development of theoretical frameworks 
\cite{lande,kk,chow} have established it as a valid conjecture in the last 
years, turning sympatric speciation into one of the favourite themes of research 
in modern evolutionary theory \cite{turelli,Gavrilets2004,Coyne2004}.

A variety of theoretical models have been proposed 
to explain sympatric speciation, from analytical mean-field type ones to  
to more realistic individual-based models. 
Computational representations based 
on variations of the Penna model for biological ageing \cite{penna}, popular amongst 
physicists working on the statistical mechanical aspects of evolutionary theory, belong 
to this latter class. Previous work on such representations have shown that 
sympatric speciation appears when driven by a change in the character of the 
distribution of ecological resources, as suggested by some biologists 
\cite{kk}. From this perspective, sympatric speciation appears as a transition 
between two different organisations of some population. In our present work, we 
develop a variation that allows a mean-field approximation with analytical 
solution, in which the nature of this transition may be further discussed.

The computational model has intra-specific competition, mutations and assortative 
mating as its sole ingredients. The mean-field approximation leads to a set of simple 
equations that reproduces some of the features of individual-based models, and 
whose solutions show a clear signature of the above mentioned phase transition.

\section{The computational model}

We take as starting point the sexual version of the Penna model, as described 
for instance in refs. \cite{revphysa,Moss99}. In addition to the age-structured 
pair of bit-strings that represents the genome for purposes of ageing analysis, each 
individual carries an extra pair of non-structured bit-strings of 32 bits each, that 
encodes a genetically 
acquired phenotype trait, as already published in ref. \cite{gustavo}. This 
extra pair of genetic material is inherited with 
the same dynamics of the age-structured pair, involving a meiotic cycle with crossing and 
recombination of each parent's bit-strings. 
The trait for a particular individual is obtained by counting the number of 
loci in the non-structured pair where the allele $1$ is either homozygous or dominant,
and is an integer in the interval $[0,32]$ which determines the 
individual's survival probability and its mating preferences. 
The positions where the allele
$1$ is dominant are chosen randomly at the beginning of the simulation and
are the same for all individuals. According to this 
number, the population is divided into three groups (subpopulations). We will 
follow the dynamical evolution of the size of the three subpopulations 
independently ($P_1$ for the one with small values of the phenotype trait, 
$P_2$ for the one with large values, and $\Pint$ for the intermediate one). The survival 
probability is $1 - V$, where $V$ is the so-called (modified) Verhulst factor. 
This factor has a resource-size parameter, the carrying capacity $C$, and 
represents a mean-field competition for the ecological resources of the 
environment. It has a different value for each one of the three subpopulations, 
representing different levels of competition for those resources: 

\begin{equation} 
     V_{P_{1,2}} = \frac{P_{1,2}+\Pint}{C}, \,\,\, 
     V_{\Pint} = \frac{x (P_1+P_2) + \Pint}{C}
\end{equation}

where we set $C=100,000$. The intermediate subpopulation $\Pint$ competes with 
a fraction $x$ of the sum 
of the subpopulations with extreme values of the phenotype trait, and this 
fraction will drive the speciation phase transition. Each of the subpopulations 
$1$ and $2$ competes only with itself and with the intermediate one. This 
variation of the Verhulst factor has previously been used in a study of 
sympatric speciation in food webs \cite{santafe}. A genetic trait, encoded by a 
single bit and subject to mutation, determines female selectivity in mating.  
This trait is initially set to zero: every female selects a mating partner 
randomly. Observe that due to mutations the offspring of a selective female may be 
non-selective and vice-versa. Mating preference also depends on the value of the 
phenotype trait. A 
selective female of population $P_1$ or $P_2$ chooses to mate, among a set of $A$ 
males from its own subpopulation, the one with the most extreme value of the phenotype trait. 
A selective female of population $\Pint$ chooses randomly to act as one of the above.
Any non-selective female mates randomly.
The number $A$ {of available males} is a measure of the female's selectivity degree: the 
larger $A$ is, more selective is the female.

\section{Results of the computational model}

\begin{figure}
\centerline{
\psfig{file=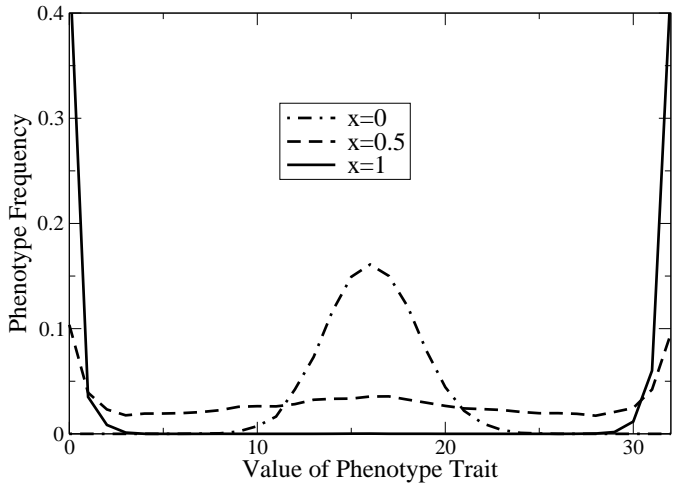,width=8cm,angle=270}}
\caption{The frequency of individuals as a function of the value of the 
         phenotype trait in the final state of the simulations, for 
         some values of the competition degree $x$.}
\label{Phen_d}
\end{figure}

\begin{figure}
\centerline{
\psfig{file=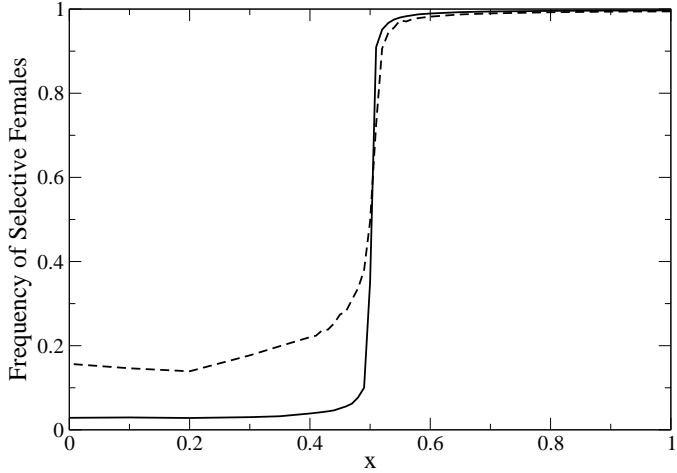,width=8cm,angle=270}}
\caption{Results of the computational model showing the fraction of selective 
         females for the values $A=50$ (solid line) and $A=3$ (dashed line) as 
         a function of $x$. Speciation occurs when this fraction reaches a 
         value close to $1$.}
\label{N_sx}
\end{figure}

We focus on the identification of the phase transition already mentioned. Fig. 
\ref{Phen_d} compares the final states of simulations carried out with extreme 
values of $x$ with the one at the transition point $x=0.5$. In
all cases, $A=50$, the mutation probability at birth of a locus of the
phenotype trait is $0.01$, and total time equals $100,000$ MC steps. Speciation is well 
observed for $x=1$, where complete reproductive isolation leads to the absence of 
gene flux between the extreme populations and consequently to the extinction of 
intermediate phenotypes. Genetic variety and gene 
flux increases crucially at the transition point $x=0.5$. The signature of the 
transition is the abrupt change of the fraction of selective females in the 
population, $N_s$, which acts as an order parameter. In a single species 
environment, this fraction is close to $0$, and it increases to $1$ with the 
event of speciation. As the control parameter $x$ is increased from $0$, the 
order parameter $N_s$ undergoes a clear transition, as seen in fig. \ref{N_sx}. 
A similar transition also occurs if a selective female chooses the mating partner 
that most closely matches her own phenotype trait, instead of the most extreme 
one \cite{karen1,karen2,karen3}, showing that the extreme mating strategy is not crucial. 

At the transition point the fluctuation of $N_s$, measured as the 
mean deviation of multiple realizations of the simulations, presents a peak. As the 
selectiveness parameter $A$ is increased, so does the steepness of the 
transition. The transition can also be seen in the behaviour of $P_1$, $\Pint$, 
$P_1 + P_2 + \Pint = P_{total}$, and $P_1 - \Pint$. In fact, we will single out 
this last quantity to signal the occurrence of speciation, as commented 
below. A full description of the nature of the sympatric speciation transition 
as obtained via the computational microscopic model is being published 
elsewhere \cite{karen1,karen2,karen3}.

\section{Mean-field approximation}

A mean-field approximation to the microscopic model can be cast under the form 
of a system of coupled differential equations. Dynamics of all three 
subpopulations are frequency-dependent; for the intermediate subpopulation, we 
add a competition for ecological resources with a fraction $x$ of each of the 
extreme-sized ones, as done in the microscopic model. The system of 
equations is thus:

\begin{equation} 
\frac{d P_1}{dt} = (a-b) P_1 + b \Pint - \frac{1}{C} (P_1+\Pint) P_1
\label{p1}
\end{equation}

\begin{equation} 
\frac{d P_2}{dt} = (a-b) P_2 + b \Pint - \frac{1}{C} (P_2+\Pint) P_2
\end{equation}

\begin{equation} 
\frac{d \Pint}{dt} = (a-2b) \Pint + b P_1 + b P_2  - \frac{1}{C} (x P_1+x P_2+\Pint) 
\Pint
\end{equation}

The parameter $a$ describes the birth rate, and is the same for all 
subpopulations. In order to characterise the exchange parameter $b$, 
we have to envision one process before as well as another one 
after crossing and recombination. 
In the first, the bits of the 
phenotypic trait are reshuffled by crossing-over of the gametes of both 
parents, a process that can lead to drastic changes in the value of this trait. 
The amount of this change is controlled by the number of males ($A$) each 
female will choose from as mating partner, as well as by the number of 
selective females in the population. Additionally, mutations may change the 
number that characterises an individual's trait. This last process is 
independent from the reshuffling and thus can be described by a constant value. 
These two processes are rather difficult to include in a mean-field model. 
For simplicity, we chose to model them by postulating that each subpopulation 
with extreme phenotype generates a fraction $b$ of its offspring with a 
phenotype of the intermediate subpopulation. The latter looses a fraction of 
$2b$ offspring to the extreme subpopulations. The parameter $b$ synthesises the 
combined effect of the mutation rate and of the degree of assortativity in the 
mating process, which should be proportional to the reciprocal value of $A$. 
Thus, even if assortative mating is maximum, the parameter $b$ cannot be set to 
zero, since we still have to model the effect of mutations. We will refer to 
$b$ as a parameter of female selectivity in the following, keeping in mind that 
it also contains the effect of mutations. Competition is introduced through the 
density-dependent Verhulst factor, to which the parameter for the carrying 
capacity $C$ is related.

Because of the intrinsic symmetry of the model, the subpopulation with a small 
value for the phenotype trait ($P_1$) has a dynamical evolution equivalent to 
the one with a high value for the trait ($P_2$). In each of these, the time 
evolution of its size depends on its current value as well as on the size of 
the intermediate subpopulation. We assume, as an approximation, that they are 
equal at all times, and set $P_1 = P_2$ in eq.~(\ref{p1}):

\begin{equation} 
\frac{d \Pint}{dt} = (a-2b) P_i + 2b P_1  - \frac{1}{C} (2x P_1+P_i) P_i.
\label{pi}
\end{equation}


The system of eqs. (\ref{p1}) and (\ref{pi}) can be simplified by the 
following transformations:

\begin{equation}
f = \frac{1}{4C}(P_1 + P_i);\,\, g = \frac{1}{4C}(P_1 - P_i);\,\, 
\epsilon = 2x - 1,
\end{equation}

$\eps$ being a control parameter in terms of which the transition is set at 
$\eps=0$. As a result of this transformation, the system of equations now reads:

\begin{equation}
\frac{df}{dt} = a f + b g - (\epsilon + 4) f^2 + \epsilon g^2,  
\label{f}
\end{equation}

\begin{equation}
\frac{dg}{dt} = (a- 3b) g - \epsilon g^2 - 4 f g + \epsilon f^2.
\end{equation}

To look for characteristics of the stationary solutions, which are the fixed 
points of the differential system, we set the time derivatives to $0$ and 
obtain a relation between the functions $g$ and $f$:

\begin{equation}
   g = -f \frac{4f - a}{4f - a + 2b}
\label{gxf}
\end{equation}

\begin{figure}
\centerline{
\psfig{file=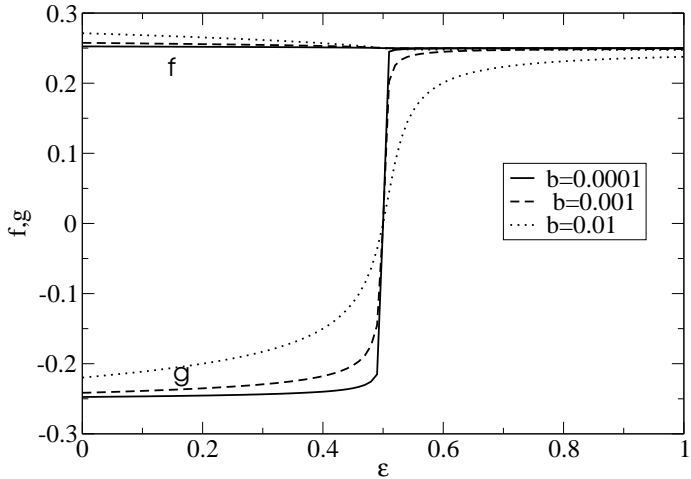,width=8cm,angle=270}}
\caption{The stable fixed points of the mean-field model for different 
         values of $b$, characterising female selectivity. 
         When $b$ increases, the transition becomes smoother.}
\label{Pepsilon}
\end{figure}

The existence of the phase transition in the mean-field approximation is 
clear in fig. \ref{Pepsilon}, in which the exact and stable solutions for the 
size of the subpopulations are shown as a function of the new control 
parameter $\epsilon$. The transition point $\epsilon = 0$ separates a region 
in phase space where the subpopulations still interbreed ($\epsilon < 0$) and 
the intermediate population is large, from another one, in which the 
subpopulation that results from interbreeding all but vanishes, characterising 
assortative mating within each of the subpopulations. In the latter, these can 
now be said to constitute two different non-mating species.

\begin{figure}
\centerline{
\psfig{file=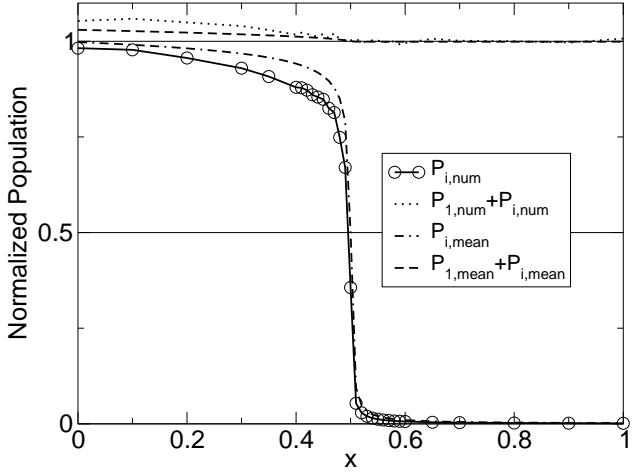,width=8cm,angle=270}}
\caption{Comparison of the result for $P_{i,num}$ of the 
         computational model ($A = 50$) with the mean-field 
         approximation $P_{i,mean}$, fitted with $b=0.001$. The 
         quantity $f \propto P_1+P_i$ is nearly constant. The horizontal
         full straight lines are guides to the eyes.}
\label{comp2d}
\end{figure}

Exploration of the parameter space shows that this transition becomes smoother 
as the selectivity degree parameter $b$ increases, as shown in fig. 
\ref{Pepsilon}. In all cases, the function $f$ is nearly constant 
(in fig. \ref{comp2d} compared to the computational model), 
as can be seen directly from the results of the 
simulations of the computational model. The latter results also show fluctuations 
in the values of all subpopulations, as well as in $g$, that peak as the 
transition point is approached. This is not true for $f$ though, for which the 
fluctuations are small and do not show any change of behaviour at the 
transition. The role played by the female selectivity $A$ is equivalent to the 
corresponding parameter in the mean-field approximation, $b$: an increase in 
selectivity $A$, corresponding to a decrease in $b$, sharpens the transition.

In order to obtain a full characterisation of the transition, and supported 
by the results of the simulations, we impose $f$ to be some constant from the 
start, and independent of $\epsilon$. Additionally, we neglect the term $bg$ in 
eq. (\ref{f}). The relations $f>0$, $f>g$ as well as $b \ll a$, valid 
in all cases we have studied, justify this simplification. It is easy to compute the 
value of $f$ from the differential system at the transition point 
$\epsilon = 0$, $f = \frac{a}{4}$. $f$ can be set constant also because of the 
following reason: If we add a small perturbation $\delta$ to $f$ in eq. 
\ref{gxf} we obtain $g = - \delta \frac{a}{2b}$ with $b \ll a$. We are also 
left with a single differential equation for $g$,

\begin{equation}
\frac{dg}{dt}=-\epsilon g^2 - 3bg + \epsilon \frac{a^2}{16}
\end{equation}

with a stationary solution that exhibits symmetry with respect to the 
transition point:

\begin{equation}
g_*= -\frac{3b}{2 \eps} +  \sqrt{(\frac{3b}{2 \eps})^2+(\frac{a}{4})^2}.
\end{equation}

\begin{figure}
\centerline{
\psfig{file=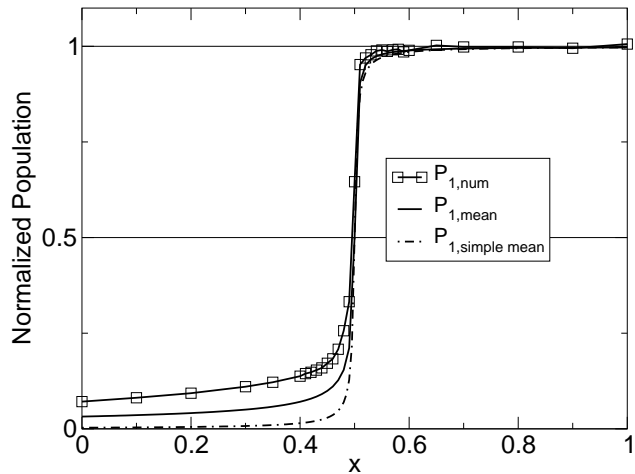,width=8cm,angle=270}}
\caption{Comparison of the result for $P_{1,num}$ of the 
         computational model ($A = 50$) with the mean-field 
         approximations $P_{1,mean}$, fitted with $b=0.001$. 
         The simplified mean-field model $P_{1,simple\, mean}$ also 
         reproduces well the simulations. }
\label{comp2dp1}
\end{figure}

To establish the validity of these last approximations, we may compare the above 
result with the stable solution of the full mean-field solution. Fig. 
\ref{comp2dp1} shows both results, and compares them to the numerical results. 
Deviations do exist, but they do not change the qualitative nature of the 
solutions.

We now proceed to obtain the time dependence of the solution. This behaviour 
could be interpreted as if, after reaching the stationary state, the value of 
the control parameter was changed; the time behaviour of the solution is then 
followed as it approaches a new equilibrium. This can be done more easily if we 
apply the transformation $z = \frac{1}{g - g_*}$ to get the equation for 
$z(t)$,

\begin{equation}
\frac{dz}{dt} = \eps + \sqrt{9b^2+\eps^2 \frac{a^2}{4}} \cdot z = \eps + rz
\end{equation}

with a time-dependent solution for $g(t)$ given by 

\begin{equation}
g(t) = -\frac{r}{\eps} \cdot 
\frac{1}{1-\phi e^{rt}} + g_*
\end{equation}

The parameter $\phi$ depends on the initial condition. If it has a value in 
the interval $0 < \phi < 1$, then $g(0)<g - 2f = -P_1 - 3P_i$, which lies 
beyond the range of $g$. The analytical solution is a very good approximation 
to the result of the simulations, yielding the same exponential behaviour. 

If we impose the condition that $f$ has a constant value, still at the 
transition point $\epsilon = 0$, then the solution for $g(t)$ is 
$g(t) = g_0 e^{-3bt}$, whereas the full solution, neglecting 
only the term $bg$ is given by

\begin{equation}
f(t) = \frac{a}{\beta e^{-at}+4} , \, g(t) = g_0 e^{-3bt} \frac{1}{\beta e^{-at}
+4}
\end{equation}

where $g_0$ and $\beta$ depend on the initial conditions. Unfortunately, the 
exponential decay we obtain in the analytical solution does not allow us to 
determine a precise relation between the selective strength parameters of the 
microscopic and mean-field models, due to too high a level of fluctuations.

\section{Conclusions}

The microscopic models for the study of sympatric speciation, and in 
particular those based on variations of the Penna model, have proven their 
value yielding a number of interesting results and providing some background 
for a testing ground of evolutionary theories. We have here reported that 
one of those versions admits a mean-field approximation with an analytical 
solution that closely matches the results of the simulations. This new tool 
can be helpful in the characterisation of sympatric speciation as an 
out-of-equilibrium phase transition, and help in the study of the statistical 
properties of the system on both sides of that transition. The thermodynamical 
nature of this transition may also be analysed in this context, as well as the 
character of the fluctuations and the quantitative properties connected to 
the transition point. Work on these lines is already in progress.

\end{document}